%% file: ppomega_mschulte_COSY-TOF_2006.tex
\documentclass{elsart}

\usepackage{epsfig}

\usepackage{amssymb}
\usepackage{stmaryrd}

\begin{document}

\begin{frontmatter}



\title{Comparison of Isoscalar Vector Meson Production Cross Sections in Proton-Proton Collisions }

\thanks[geldgebern]{Supported by BMBF (Germany) and the FZ-J\"ulich.}
\begin{center}
\large
The COSY-TOF collaboration
\normalsize
\end{center}

\author[juel]{M. Abdel-Bary}
\author[juel]{S. Abdel-Samad}
\author[tudd]{K.-Th. Brinkmann}
\author[tueb]{H. Clement}    
\author[tudd]{J. Dietrich}
\author[tueb]{E. Doroshkevich}
\author[ross]{S. Dshemuchadse}   
\author[tueb]{K. Ehrhardt}    
\author[tueb]{A. Erhardt}    
\author[erla]{W. Eyrich}    
\author[tori]{A. Filippi}	
\author[tudd]{H. Freiesleben}
\author[erla]{M. Fritsch}    
\author[juel]{A. Gillitzer}		 
\author[juel]{D. Hesselbarth}
\author[tudd]{R. J\"akel}
\author[tudd]{L. Karsch}
\author[juel]{K. Kilian}  	 
\author[tudd]{E. Kuhlmann}
\author[tori]{S. Marcello}   
\author[ross]{P. Michel}
\author[ross]{K. M\"oller}   
\author[juel]{H.P. Morsch}		
\author[erla]{C. Pizzolotto}    
\author[tudd]{Ch. Plettner}
\author[juel]{J. Ritman} 	   
\author[juel]{E. Roderburg} 	   
\author[erla]{P. Sch\"onmeier}
\author[erla]{W. Schroeder}    
\author[tudd]{M. Schulte-Wissermann\corauthref{corauthlabel}}
\ead{M.Schulte-Wissermann@physik.tu-dresden.de}
\author[boch]{M. Steinke}
\author[tudd]{G.J. Sun}
\author[tudd]{W. Ullrich} 
\author[tudd]{R. Wenzel} 
\author[juel]{P. Wintz}       
\author[erla]{M. Wagner}     
\author[boch]{A. Wilms} 
\author[erla]{S. Wirth}
\author[wars]{P. Zupranski}

\corauth[corauthlabel]{Corresponding Author}
\address[tudd]{Institut f\"ur Kern- und Teilchenphysik, Technische Universit\"at Dresden, D-01062 Dresden, Germany}
\address[boch]{Institut f\"ur Experimentalphysik, Ruhr-Universit\"at Bochum, D-44780 Bochum, Germany}
\address[erla]{Physikalisches Institut, Universit\"at Erlangen-N\"urnberg,D-91058 Erlangen, Germany}
\address[juel]{Institut f\"ur Kernphysik, Forschungszentrum J\"ulich, D-52425 J\"ulich, Germany}
\address[ross]{Institut f\"ur Kern- und Hadronenphysik, Forschungszentrum Rossendorf, D-01314 Dresden, Germany}
\address[tori]{University of Torino and INFN, Sezione di Torino, I-10125 Torino, Italy}
\address[tueb]{Physikalisches Institut, Universit\"at T\"ubingen, D-72076 T\"ubingen, Germany}
\address[wars]{Soltan Institute for Nuclear Studies, 05-400 Swierk/Otwock, Poland}

\begin{abstract}
The reaction $ pp\to pp\bf \omega$ was investigated with the TOF spectrometer, which is 
an external experiment at the accelerator COSY (Forschungszentrum J\"ulich, Germany). 
Total as well as differential cross sections were determined at an excess energy of $93\, MeV$ ($p_{beam}=2950\,MeV/c$). 
Using the total cross section of $(9.0\pm 0.7 \pm1.1)\,\mu b$ for the reaction $ pp\to pp\omega$ determined here 
and existing data for the reaction $pp\to pp\bf \phi$, the ratio $\mathcal{R}_{\phi/\omega}=\sigma_\phi/\sigma_\omega$
turns out to be significantly larger than expected by the Okubo-Zweig-Iizuka (OZI) rule. The 
uncertainty of this ratio is considerably smaller than in previous determinations.
The differential distributions show that the $\omega$ production is
still dominated by S-wave production at this excess energy, however  higher partial waves clearly contribute. 
A comparison of the measured angular distributions for $\omega$ production to published
distributions for $\phi$ production at $83\,MeV$ shows that the data are consistent with
an identical production mechanism for both vector mesons.
\end{abstract}

\end{frontmatter}

\section*{Introduction}
Near-threshold production of the isoscalar vector mesons, $\omega$ and $\phi$,  
in proton-proton interactions 
remained largely unstudied until the late 1990s. 
Then, first experimental data for the $\omega$-production directly at threshold \cite{Hibou99} and 
for both mesons at moderate excess energies ($\epsilon < 320\, MeV$) became available \cite{AbdEl-Samad:2001dv,Balestra:2001}.
In parallel, a considerable interest from theory arose 
\cite{cite:hanhart98,Titov00,Tsushima03,Fuchs03,FaesslerFuchs03,cite:faesslerfuchs04}, addressing the question of 
reaction dynamics, of possible proton-vector meson resonances, 
and of in-medium effects of vector meson properties. 
Another important question is a possible $s\overline{s}$ content of the nucleon wave function, which may be determined 
through the ratio of the total cross sections of
$\omega$ and $\phi$ mesons in proton-proton collisions at identical excess energies
($\mathcal{R}_{\phi/\omega}=\sigma_{pp\to pp\phi}/\sigma_{pp\to pp\omega}$).
This comes about since the  flavor eigenstates ($\omega_0, \omega_8$) 
of the vector nonet are arranged in such a way that the mass eigenstates ($\phi, \omega$)
form a quasi-ideally decoupled system 
($|\phi\rangle \cong |s\overline{s}\rangle, |\omega\rangle \cong |u\overline{u}\rangle 
+ |d\overline{d}\rangle)$, where
the small deviation of $3.7^\circ$ \cite{cite:pdgmitwinkel} from the ``ideal mixing angle'' of $35.3^\circ$ 
yields the  $|u\overline{u}\rangle + |d\overline{d}\rangle$ admixture to the $\phi$ wave function.
According to the Okubo-Zweig-Iizuka (OZI) rule \cite{OZIcomp} processes involving disconnected
quark lines are \emph{strongly} suppressed, 
so that the production of $\phi$ mesons can  take place only via the small admixture 
of \emph{non-strange} quarks.
Based on the deviation from the ``ideal mixing angle'' Lipkin \cite{cite:lipkin76} predicted the ratio
of the production cross sections of $\phi$ to $\omega$ mesons 
to be $\mathcal{R}_{OZI} = 4.2\times10^{-3}$. However, $\mathcal{R}_{OZI}$ is exceeded experimentally in many independent 
determinations.
This fact is often denoted as \emph{``violation of the OZI rule''}.
Using data from proton-proton interactions \cite{Hibou99,AbdEl-Samad:2001dv,Balestra:2001,cite:anke06} 
 $\mathcal{R}_{\phi/\omega}\approx 8\times \mathcal{R}_{OZI}$ is found at
excess energies $\epsilon < 100\, MeV$, where the combined uncertainties calculated from the statistical and systematic 
errors range from about $30\%
$ to more than $50\%
$.
In contrast to the data close to threshold, 
at high excess energies \mbox{($\epsilon>1\,GeV$)} only $(1-2.4) \times \mathcal{R}_{OZI}$ 
is found \cite{ABaldini98,RBaldi77,VBlobel75}. 
In $\overline{p}p$ annihilation the enhancement depends on the
momentum transfer and can be as large as a factor of 260 \cite{Alessanda00}, while
in  $\pi N$ interactions a $\phi/\omega$ enhancement of $(3.2\pm 0.8)$ was extracted \cite{cite:sibirtsev00}.\newline
From the theoretical side, the issue of ``hidden strangeness'' is controversially discussed, and 
some theoretical approaches introducing ``off-shell'' mesons \cite{Fuchs03},
higher order rescattering processes, 
and double-hairpin diagrams \cite{cite:kaptari05} succeed 
in describing a moderate enhancement of \mbox{$\mathcal{R}_{\phi/\omega}$} over $\mathcal{R}_{OZI}$. 
In addition, the initial-state-interaction could potentially influence the cross section ratio $\mathcal{R}_{\phi/\omega}$, 
since 
- due to the mass difference of $m_\phi-m_\omega = 237\,MeV$ - 
different energies in the initial state are needed in order to reach the same excess energy in both exit channels 
\cite {cite:hanhartnakajama99}.  \newline
It must be emphasized that a meaningful comparison of total cross sections of $\phi$ to $\omega$ production 
in view of ``OZI-violation''
necessarily requires the same production processes for both vector mesons.
Therefore, prior to final conclusions from measured $\mathcal{R}_{\phi/\omega}$
values, differential cross sections of both mesons have to be measured.
For proton-proton interactions no differential data exist below $\epsilon = 173\,MeV$ for $\omega$ production, while
for the $\phi$ the only differential data available are at an excess energy $\epsilon = 83\,MeV$ \cite{Balestra:2001} 
and at $\epsilon = 18.5\,MeV$ \cite{cite:anke06}.
For the lower excess energy, the $\phi$ production is found to be described by pure S-wave, 
with a sizable contribution from final-state interaction in the pp-system \cite{cite:anke06}.
 At $\epsilon = 83\,MeV$, the DISTO experiment has found that higher angular momenta contribute significantly. 
This paper will report differential cross sections for the reaction  $pp\to pp\bf \omega$ at $\epsilon =  93\, MeV$,
i.e.~only $10\,MeV$ above the DISTO measurement.
Since the involved matrix elements can be assumed to be nearly constant within this small range of excess energy, 
a direct comparison of the $\omega$ to the $\phi$ differential distributions is now possible.

\section*{Experimental Methods and Results}
\begin{figure}[ttt]
\begin{center}     
	\mbox{\epsfig{file=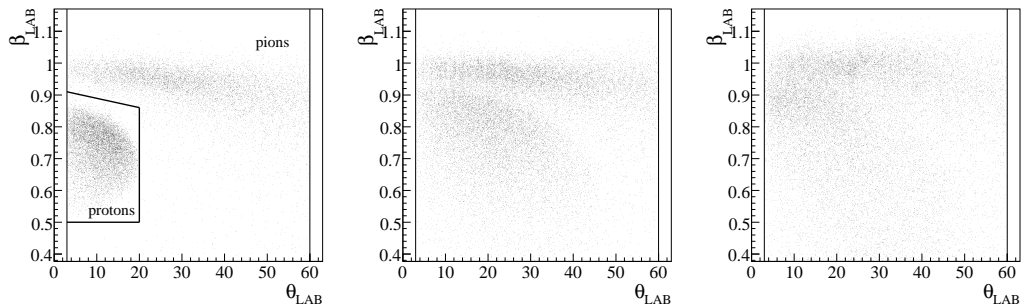,width=1.0\textwidth}}
	\caption{ \it \label{fig:thetaVsBeta}
	Left Frame: Monte Carlo distribution of protons and charged pions of the reaction
    $pp\to pp\omega,\, \omega\to\pi^+\pi^-\pi^0$ shown in a \mbox{$\beta$ vs.~$\theta$}
     plot.
	Protons and pions can clearly be separated, as the protons are restricted to the \mbox{$\beta$ vs.~$\theta$}
	region indicated by the selection box. The vertical lines show the limits of optimum acceptance of the detector.
	Center frame: Monte Carlo distribution for $pp\to pp\pi^+\pi^-$.
	The protons are less restricted 
	kinematically, which leads to a less accurate determination of the missing mass.
	Right Frame: Experimental data; totally dominated by the background.
	\rm  }
\end{center} 
\end{figure}

The Time-Of-Flight spectrometer TOF \cite{cite:tofbeschreibung} is an external experiment at the 
COoler SYnchrotron COSY (J\"ulich).
The proton beam hits a $4\,mm$ thick liquid hydrogen target
and the emerging reaction pro\-ducts traverse a layered time-of-flight start and tracking detector.
After a flight path of $\approx 3\,m$ through vacuum the 
ejectiles are detected in the highly granulated stop components of the spectrometer.
From time and position measurements the 
velocity vectors of all charged particles
are determined with a time-of-flight resolution of better than $\sigma_{TOF} = 300\, ps$ 
and angular track-resolution of better than $\sigma_{\sphericalangle}=0.3^\circ$. 
Due to the low mass area density of all detector components, 
the influence of small angle scattering and energy loss is almost negligible for
particles with $\beta>0.5$.
Only particles in this velocity range are produced in the reaction under study.\newline
Unlike magnetic spectrometers, which provide particle identification by often paying the cost
of limited acceptance, the TOF detector covers the full kinematical range ($0^\circ\le\phi<360^\circ, 
3^\circ<\vartheta<60^\circ$)
of most reactions and measures the velocity vectors of all charged particles.
Different reaction channels (e.g., $pp\to pp, d\pi^+, pK^+\Lambda, pK^+\Sigma^0$) can be identified 
unambiguously by examination of their event topology. For this, mass hypotheses are applied to the measured velocity vectors
in order to calculate the
four-momenta of the tracks. From these, a missing mass spectrum can be constructed.
Since the momentum calculation diverges as $\beta\to 1$, only tracks with velocities below $\beta\approx0.9$ 
can be used in order to determine meaningful missing mass values.\newline
In the case of $pp\to pp\omega, \omega\to\pi^+\pi^-\pi^0$ the fact is  exploited that
protons and charged pions populate disjoint kinematical regions in the $\beta$ vs.~$\theta_{LAB}$ plane \cite{MSWdiss}. 
This is shown in Fig.~\ref{fig:thetaVsBeta} in the left frame for Monte Carlo data. 
While protons are restricted to forward angles and moderate velocities,  
the pions cluster at $\beta>0.9$ over the full angular range. \newline
In a first step of the analysis, only four-prong events with two entries inside ($\to protons$) and two hits outside ($\to pions$) 
of the selection box indicated in Fig.~\ref{fig:thetaVsBeta} are 
treated as $pp\to pp\omega$ candidates. 
Using Monte Carlo simulations the assignment of protons and pions to the respective 
tracks is found to be correct for over $99.5\%
$ of all events. 
In the experiment, however, the $\omega$ signal will be hidden in a huge background of resonant and non-resonant 
multi-pion production ($pp\to ppX, X=\pi^+\pi^-, \pi^+\pi^-\pi^0, \eta\to\pi^+\pi^-\pi^0$).
In these cases the minimum invariant mass of the pion systems is smaller than the $\omega$ mass ($m_X<m_{\omega}$), 
hence the ejectiles are kinematically less restricted and the proton velocities are more elevated.
This can be seen in the middle frame of Fig.~\ref{fig:thetaVsBeta}, where simulated $pp\to pp\pi^+\pi^-$ data is shown. 
A similar picture holds for the channel $pp\to pp\eta, \eta\to \pi^+\pi^-\pi^0$. 
In multi-pion production pions can be mistaken as protons if they are found inside the selection box, and vice versa.
In addition, the higher velocities of the ``identified protons'' considerably decrease the missing mass resolution.
The misinterpretation and the high velocities lead to an incorrect missing mass which gives rise to a 
structureless and continuous background in the missing mass distribution.
The two- and three-pion channels dominate the four-prong events as can be seen in the right frame of Fig.~\ref{fig:thetaVsBeta}, 
where experimental data is shown. No indication of an $\omega$ signal can be seen, as it is swamped 
by the other reactions.\newline
In the second step of the analysis, the two-pion part of the background can be reduced by
selecting the $\omega$-decay into three pions ($\pi^+\pi^-\pi^0$, $\mathcal{BR}=0.89$). 
Here, the plane defined by the two charged pions will, in general, not contain the $pp$
missing momentum vector ($=\omega$ momentum vector) due to the momentum of the undetected $\pi^0$.
Applying an acoplanarity cut 
of $\alpha = \angle ((\overrightarrow{p}_{\pi_1}\times\overrightarrow{p}_{\pi_2}),(\overrightarrow{p}_{\pi_1}\times\overrightarrow{p}_{\omega}))>5^\circ$ 
suppresses 90\% of two-pion background,  
while only  $17\%
$ of the $\omega$ events are lost according to MC studies.
The value $\alpha=5^\circ$ was determined by optimizing the signal-to-background ratio for experimental data.
Varying $\alpha$ in the range of 1 to 10 degrees leaves the total cross section constant within $\pm2.1\%
$. \newline
Finally, only events with the combined momentum of the protons pointing into the backward hemisphere 
of the CMS are considered.
In this case the protons have smaller velocities in the laboratory frame, which improves momentum resolution. 
Apart from reducing the number of events by about $30\%
$, this restriction does not lead to any loss of phase space coverage due to the symmetric entrance channel. 

\begin{figure}[ttt]
\begin{center}     
	\begin{minipage}[t]{0.55\textwidth}
	\mbox{\epsfig{file=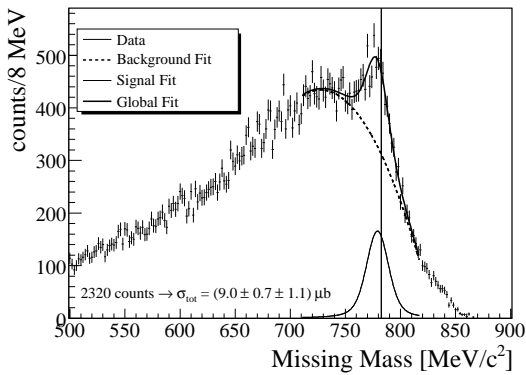,width=1.0\textwidth}}
	\end{minipage}
	\hfill \mbox \hfill
	\raisebox{5.5cm}{
	\begin{minipage}[t]{0.42\textwidth}
	\caption{ \it \label{fig:mmass}
	Missing mass spectrum calculated  from both proton tracks after applying the cuts described in the text. 
	The $\omega$ signal is clearly seen
	above a smooth multi-pion background. The signal is fitted with a Voigt function, whose 
	integral represents the total number of counts. After acceptance correction and absolute
	normalization the total cross section results to  $(9.0\pm 0.7 \pm1.1)\,\mu b$.
	\rm  }
	\end{minipage}
	}

\end{center} 
\end{figure}

Fig.~\ref{fig:mmass} shows the missing mass distribution which is obtained from the two identified protons.
A peak at the $\omega$  mass 
can be seen above a smooth multi-pion background. 
At high masses the shape of the missing-mass spectrum is governed by phase-space, while towards lower masses
the cuts described above lead to a continuous reduction.
The total number  of counts in the $\omega$ signal is obtained by a simultaneous fit of a 
second order polynomial 
and a Voigt function (convolution of a Gaussian with a Breit-Wigner function). 
The width of the Breit-Wigner distribution 
is fixed to the natural width of the $\omega$-meson ($\Gamma=8.49\,MeV$)\cite{cite:pdg}, while all other parameters 
are allowed to vary freely. 
The width of the resulting Gaussian ($\sigma=(7.7\pm0.9)\,MeV$) reflects the detector resolution and 
is in good agreement with the Monte-Carlo simulation ($\sigma=6.7\,MeV$).
The fit yields a total number of $2320$ counts and a statistical uncertainty of $\pm7.7\%
$.
Using the error of the fit as a measure for the statistical uncertainty is conservative, 
since it
incorporates the combined statistical fluctuations of signal and background. \newline
%
The detector acceptance is corrected using Monte-Carlo methods. 
Due to the large phase-space coverage of the detector, it is quite high (on average $40\%
$) for the considered reaction and varies smoothly over the full kinematical range. 
The event generator used produces a three-body phase-space distribution (5 DOF), 
where the width of the $\omega$ meson and the 
Zemach prescription of the $\omega$ decay is included ($J^P=1^-$) \cite{cite:zemachComp}. 
Differential distributions and intermediate
resonant states can be accounted for, however they have been omitted since the final state is
found to be mainly isotropic.
The produced particles are propagated through a full representation of the detector, including energy loss, hadronic interaction,
secondary particles, signal generation and final digitized output.
The Monte Carlo output is then subjected to the very same analysis
routines (including the cut settings) as the measured data, hence possible software inefficiencies are also
accounted for.
The total systematic uncertainty of the fitting process and the acceptance correction 
is determined by varying the fit ranges and the values of the above mentioned 
cuts within reasonable limits, from which a systematic uncertainty of $11.6\%
$ was deduced.
This value also includes the influence of the background shape, since the variation
of the fit ranges leads to a change of the background polynomial from a convex shape (wide background interval) to
almost a straight line (narrow background interval). \newline
The absolute normalization is accomplished at TOF by evaluating elastic scattering, which is 
simultaneously measured during the experiment. The obtained angular distribution is compared
to literature data \cite{AltmeierEDDA}, where the normalization factor directly yields the luminosity of $(1.45 \pm 0.06)/nb$. 
The  uncertainty of this procedure ($3.8\%
$) is in equal parts due to the intrinsic uncertainty of our measurement 
and the error of the literature data.\newline
 After acceptance correction and absolute
normalization a total cross section of 
$(9.0\pm 0.7 \pm1.1)\,\mu b$
 is obtained. 
 It is in agreement with our result of $(7.5\pm 1.5 \pm1.9)\,\mu b$ published in Ref.~\cite{AbdEl-Samad:2001dv},
however with improved accuracy.

\setlength{\unitlength}{1cm}
\begin{figure}[ttt]
 \begin{center}
	\mbox{\epsfig{file=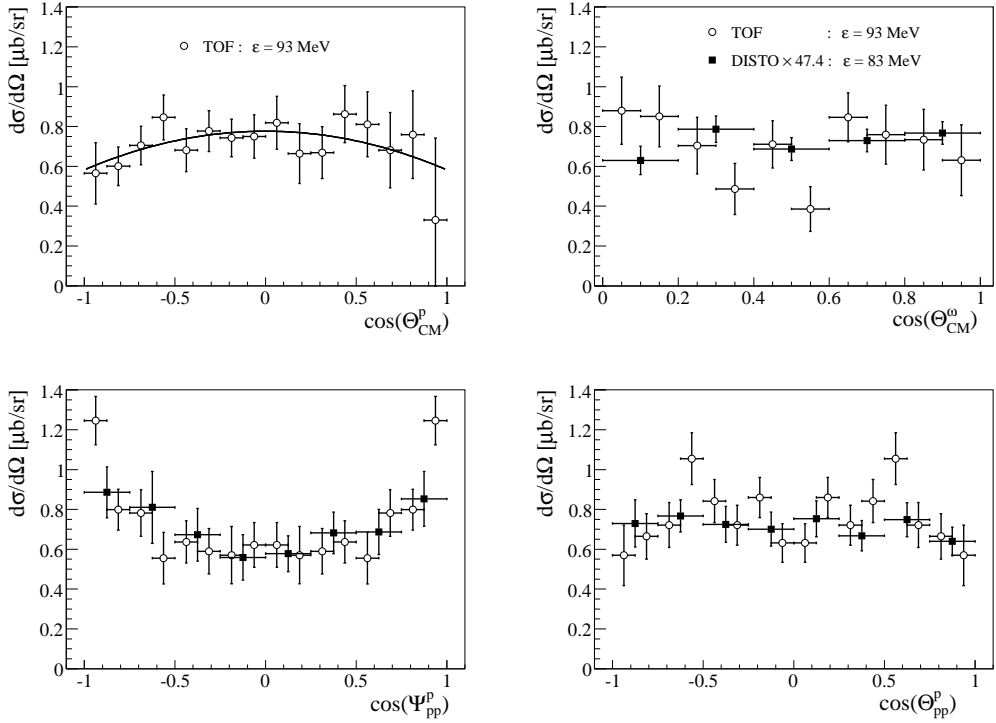,width=1.0\textwidth}}
	\caption{ \it 
			Differential data of this experiment for $pp\to pp\omega$ (open circles),
            compared to data of DISTO for $pp\to pp\phi$ \cite{Balestra:2001},
            if available (full squares).
			The error bars of the TOF data show the uncertainty
            of the fit, reflecting the statistical errors of signal and background.
		 	The DISTO data are scaled by the cross section ratio $\mathcal{R}_{\omega/\phi} = 47.4$. The variables are
			named in accordance with \cite{Balestra:2001}.
			Upper Left: differential cross section as function of $cos\Theta_{CM}^p$, i.e.~cosine of the polar angle of the protons
			in the overall CMS. 
			Upper Right: 
			angular distribution of the meson in the overall CMS.
			Lower Left:  symmetrized differential cross section as function of the angle of the meson with respect to the protons,
			measured in the final state pp rest-frame (helicity angle). Lower Right: symmetrized differential cross 
			section as function of 
		    the angle of the reaction protons with respect to the incident protons, 
			measured in the final state pp rest-frame (Jackson angle).
			\rm  
			  }
	\label{fig:ergebnisse}
 \end{center}
\end{figure}

In order to extract differential cross sections, the spectrum in Fig.~\ref{fig:mmass} is created for consecutive bins of 
the quantity under study. Each missing mass distribution is then individually fitted, integrated, and corrected for acceptance. 
The reduced number of entries for each bin 
leads to an elevated uncertainty of the fitting process, as  
the influence of the statistical fluctuations of the background becomes more important. In addition, the shape
of the background changes along with the observable under study which introduces additional uncertainties.
To minimize these effects, the widths of the Gaussians are fixed to the individual values obtained in the 
corresponding Monte-Carlo simulation.
While in most cases the fits yield uncertainties of about $18\% 
$, for some particular bins the error can reach $40\% 
$.
Fig.~\ref{fig:ergebnisse} shows the resulting angular distributions.
In the upper left frame the angular distribution of the individual protons in the overall CMS is plotted, while
in the upper right frame
the meson distribution in the overall CMS is shown. In the lower part we find the distributions of the 
helicity angle (left frame) and of the Jackson angle (right frame)\footnote{\label{page:GWerklaerung} We are using the Gottfried-Jackson frame \cite{cite:goodoljackson} 
with the two reaction protons 
residing in their common rest frame.  Following
the conventions in Ref.~\cite{cite:byckling,cite:sibirtsevjackson},
the momenta of the reaction $pp\to pp\,\omega$ are labeled as $\overrightarrow{p}_a \overrightarrow{p}_b\to \overrightarrow{p}_2 \overrightarrow{p}_3\, \overrightarrow{p}_1$,
hence the rest-frame under consideration (reaction protons) is the \{2,3\} sub-system.
While in a ``normal'' CM or LAB system  $\overrightarrow{p}_a + \overrightarrow{p}_b = \overrightarrow{p}_1 + 
\overrightarrow{p}_2 + \overrightarrow{p}_{3}$ 
holds, in the GJ-frame we have 
$\overrightarrow{p}_2 + \overrightarrow{p}_3 = \overrightarrow{p}_a + \overrightarrow{p}_b - \overrightarrow{p}_{1} = 0$.
In this Lorentz frame 
the helicity angle is defined as $\sphericalangle(\overrightarrow{p}_3,\overrightarrow{p}_1)$, while
the Jackson angle is defined as $\sphericalangle(\overrightarrow{p}_3,\overrightarrow{p}_b)$. 
}.\newline
The DISTO collaboration published differential distributions of various variables for the reaction 
$pp\to pp\phi$ at $\epsilon=83\,MeV$ \cite{Balestra:2001}.
Three of these distributions are plotted in the corresponding frames in Fig.~\ref{fig:ergebnisse}.
In order to facilitate a direct comparison of $\omega$ and $\phi$ production the DISTO data 
are scaled by the cross section ratio ($\sigma_\omega/\sigma_\phi = 47.4)$.
The $\omega$ and $\phi$ distributions are in general agreement with each other, which can be seen in
the results of a Legendre fit using the first two 
even coefficients (Tab.~\ref{tab:legendreresultsV2}).\newline
Several consistency checks can be made as stringent quality tests of the four differential distributions
presented here. 
Firstly, 
the background parameters vary smoothly and continously over the full angular range in all differential 
distributions.
Secondly, the integration of each differential distribution shown must result in the total cross section.
Using $\sigma_{\omega}=4\pi\cdot a_0$ and the values given in Tab.~\ref{tab:legendreresultsV2} this is indeed the case for all 
distributions within $5\%
$.
Finally, one can check the reflection symmetry of the exit channel, which 
follows from the symmetric entrance channel.
This is shown in Fig.~\ref{fig:ergebnisse} in the upper left frame,
where the unsymmetrized angular distribution of the individual protons in the overall CMS is plotted.
It is indeed symmetric about $cos\theta=0$ (a fit including 
also the first odd Legendre polynomial yields 
$a_0= (0.72\pm0.04)\,\mu b/sr$, $a_1= (0.02\pm0.07)\,\mu b/sr$, and $a_2= (-0.12\pm0.09)\,\mu b/sr$).
In addition, this distribution shows that the acceptance correction is well under control, since its influence
is high in this particular case: Due to the requirement to accept only events with the 
combined $pp$-momentum pointing into the backward CMS hemisphere ($cos\theta_{pp}<0$) the probability of single protons 
to emerge with $cos\theta_p>0$ is reduced (the acceptance drops from $>60\%
$ at \mbox{$cos\theta_p\approx -1$} to $7.2\%
$ at \mbox{$cos\theta_p\approx 1$}).
For all other differential observables shown in Fig.~\ref{fig:ergebnisse} the acceptance 
is constant within $\pm20\,\%
$.

\section*{Discussion}
\subsection*{Total Cross Sections}
Assuming absolute OZI suppression of processes with disconnected quark lines and
identical production mechanisms for $\omega$ and $\phi$ mesons, the 
deviation of the cross section ratio $\mathcal{R}_{\phi/\omega}$ from
$\mathcal{R}_{OZI}$ is a measure for the strangeness content in the nucleon:
\begin{equation}
\mathcal{R}_{\phi/\omega}
= \frac{\sigma_{pp\to pp\phi}}{\sigma_{pp\to pp\omega}}
= c\times tan^2\left(\Delta\theta_V\right) = c\times4.2\cdot10^{-3} =  c\times\mathcal{R}_{OZI},
\label{eqn:Rozisimple}
\end{equation}
where $c$ 
para\-metri\-zes the strength of the OZI-violation and
$\Delta\theta_V=3.7^\circ$ is the deviation from the ideal mixing angle.
In addition to eq.~(\ref{eqn:Rozisimple}), Lipkin predicted 
that the ratio of the $\phi\rho\pi$ and $\omega\rho\pi$ coupling constants ($g^2_{\phi\rho\pi}/g^2_{\omega\rho\pi}$)
should also
yield $tan^2(\Delta\theta_{V})=\mathcal{R}_{OZI}$ \cite{cite:lipkin76}. 
From experimental data a ratio is deduced which exceeds this prediction 
by a factor of three \cite{cite:sibirtisev05}, i.e. the  ($g^2_{\phi\rho\pi}/g^2_{\omega\rho\pi}$) ratio
itself violates the OZI-rule. 
This fact implies, however,
that theoretical models with the dominant production mechanism  for $pp\to pp\phi$ and $pp\to pp\omega$ given by
these coupling constants 
consequently succeed in a description of $\mathcal{R}_{\phi/\omega}\approx3\times \mathcal{R}_{OZI}$.
Therefore, only enhancement factors significantly larger than $c=3$ will add new ``exotic'' 
ingredients to the OZI-puzzle.

Recently, ANKE measured ${\sigma}_{pp\to pp\phi}$ 
at an excess energy of $75.9\,MeV$ \cite{cite:anke06}.
The experimental value for the total cross section is $\sigma_{pp\to pp\phi}=(188.0\pm19.1\pm41.4)\,nb$, where the 
experimental uncertainty is roughly a factor of 2 smaller than in Ref.~\cite{Balestra:2001}.
The overall uncertainty of $\sigma_{pp\to pp\omega}$ presented here is only about half
of the value published in Ref.~\cite{AbdEl-Samad:2001dv}, so that $\mathcal{R}_{\phi/\omega}$
can now be calculated with a smaller uncertainty.
For a comparison of the total cross sections we have to account for the difference 
in excess energy ($\epsilon=93\,MeV \rightarrow \epsilon=75.9\,MeV$). 
For this extrapolation we use the parameterization of Sibirtsev \cite{Sibirtsev96}, 
which describes the world data remarkably well from $\epsilon>1\,GeV$ down to the TOF energy region:
\begin{equation}
\sigma_{pp\to pp\omega} = a\left(1-\frac{s_0}{s}\right)^b \left(\frac{s_0}{s}\right)^c.
\label{eqn:sibirtsevsparamet}
\end{equation}
Here, $s=(2m_p+m_{\omega}+\epsilon)^2$ is the squared invariant mass of the total system and $s_0=(2m_p+m_{\omega})^2$ 
is the threshold value. 
The parameters $b$ and $c$ are taken from Ref.~\cite{Sibirtsev96} ($b=2.3,
c=2.4$). The parameter $a=5.3\,\mu b$ 
is fixed to reproduce our measured cross section at $\epsilon=93\,MeV$.
Then, eq.~(\ref{eqn:sibirtsevsparamet}) yields a total cross section of 
$6.0\,\mu b$ at $\epsilon=75.9\,MeV$, 
which leads to the experimental value 
$\mathcal{R}_{\phi/\omega} = (31\pm 8) \times10^{-3}$, or $(7.5\pm2.1) \times\mathcal{R}_{OZI}$.
Considering the uncertainty, which is dominated by the $\phi$ cross section, 
the enhancement factor is neither unity (``na\"ive OZI'') nor does it agree with 
 $g^2_{\phi\rho\pi}/g^2_{\omega\rho\pi}\approx3\times\mathcal{R}_{OZI}$. 
This experimental finding is also supported by measurements at smaller excess energies. Using the data of 
Ref.~\cite{cite:anke06} for $\sigma_\phi$ and of Ref.~\cite{Hibou99} for $\sigma_\omega$, 
one finds an enhancement factor of $10.1$ and $7.7$ for 
$\epsilon=34.5\,MeV$ and $\epsilon=18.5\,MeV$, respectively, both with an uncertainty of roughly  $30\%
$.
This leads to the conclusion that $\mathcal{R}_{\phi/\omega}$ is considerably larger than $\mathcal{R}_{OZI}$ starting 
from near the threshold up to $\epsilon\approx100\,MeV$.
On the other hand, at excess energies larger than $1\,GeV$ only 
$\mathcal{R}_{\phi/\omega}\approx (1-2.4) \times \mathcal{R}_{OZI}$ is found. 
Whether the enhancement of $\mathcal{R}_{\phi/\omega}$ over $\mathcal{R}_{OZI}$ near threshold is a sign for 
``hidden strangeness'' in the nucleon, or due to the initial- or final-state-interaction,  
or a dynamical effect hidden in the production processes, 
or even an indication for a cryptoexotic resonance in the $p\phi$ system as suggested in  Ref.~\cite{cite:sibirtisev05} - 
all this is far from clarification. 

\subsection*{Differential Cross Sections}
The three-body final state ($pp\omega$) can be described using two angular momenta:
(1) the orbital angular momentum  of the two protons relative to each other ($l_1$), 
and (2) the orbital momentum of the $\omega$  relative to the proton-proton system ($l_2$). 
At threshold, both angular momenta have to be zero ($l_1=l_2=0$), and 
due to parity and angular momentum conservation, the entrance channel will be a $^3P_1$ state.
This has been verified experimentally in Ref.~\cite{cite:anke06} for the reaction $pp\to pp\phi$ at $\epsilon=18.5\,MeV$. 

\input{ppomega_mschulte_COSY-TOF_2006_Tab1}

At $\epsilon\approx90\,MeV$ the angular distribution of the vector mesons in the overall CMS 
(upper-right frame of Fig.~\ref{fig:ergebnisse}) is isotropic within the experimental uncertainty. 
An isotropic distribution is a necessary condition for the angular momentum  
$l_2$ between the $pp$-system and the $\phi/\omega$ meson to be zero. 
From the consistency with an isotropic distribution alone, however, it cannot be
concluded that the angular momentum $l_2$ between the $pp$-system and the $\phi/\omega$
meson must be zero, since cancellation effects of higher partial waves may also result
in an isotropic distribution.
It should be mentioned that $l_2=0$ 
is somewhat surprising since the maximum momentum of the vector meson in the CMS is $p^{*}_{max} \approx 330\,MeV/c$, 
hence contributions from higher partial waves should be possible. 
If, however, $l_2=0$ holds, only $J^P=\frac{1}{2}^-$ nucleon resonances 
can contribute to $\omega$ production via $pp\to pN^*, N^*\to p\omega$.\newline
In contrast to this isotropic distribution, the 
 angular distribution of the $\phi$  shows a significant anisotropy 
when measured in the final state  proton-proton rest frame 
(helicity angle, lower-left frame of Fig.~\ref{fig:ergebnisse}).
Within uncertainty, the same is found for $\omega$ production, as can be seen in \mbox{Tab.~\ref{tab:legendreresultsV2}} 
where the numerical values of the Legendre fits are summarized.
A non-isotropic angular distribution is only possible if partial waves higher than S-waves contribute.
DISTO computes $|M_{10}|^2/(|M_{00}|^2+|M_{10}|^2)=0.28\pm0.07$ for the 
ratio of the matrix elements with $l_1=1,\, l_2=0$ and $l_1=l_2=0$ \cite{Balestra:2001}, however, without explicitly ruling out higher
partial wave contributions\footnote{Note that according to chapter 4.1.1. of Ref.~\cite{cite:hanhartreport} any non-isotropic distribution of the helicity angle 
 requires $l_1>0$ and $l_2>0$.}.
Since the $a_2/a_0$ ratio of both reactions is comparable within uncertainties,
one can assume a similar value for the ratio of the matrix elements governing $\omega$ production. Hence,
the vector meson production at $\epsilon\approx90\,MeV$  still mainly proceeds through $l_1=0,\, l_2=0$, however,
in this energy region we see the onset of higher angular momenta. 
 As a consequence, the initial partial wave
is no longer necessarily $^3P_1$ as at threshold.\newline
The last frame (lower-right in Fig.~\ref{fig:ergebnisse}) 
shows the distribution of the Jackson angle, i.e.~the angular distribution of the incident protons
measured in the final state  proton-proton
reference frame (see footnote on page \pageref{page:GWerklaerung}).
While the DISTO distribution for the $\phi$ is isotropic, the 
$\omega$ distribution shows some additional structure. In fact, a fit with the first three even Legendre 
polynomials yields $a_0=(0.73\pm0.04)\mu b/sr$, $a_2=(-0.22\pm0.15)\mu b/sr$, $a_4=(-0.39\pm0.20)\mu b/sr$
which are only poorly compatible with isotropy. 
However, considering the statistical and systematical uncertainties described above, 
an isotropic distribution cannot be ruled out by our data. 

The angular distributions presented here for the  $\omega$ production in proton-proton collisions
agree in shape with those for $\phi$ production.
Thus, regarding OZI-violation, the \emph{assumption} of 
similar production mechanisms for both mesons is  valid, at least for this excess energy.
However, it must be emphasized that the data by no means prove identical dynamics for both systems.
For example, in the meson exchange model of Nakayama and Tsushima \cite{Tsushima03} 
the individual contributions of mesonic, nucleonic, and resonant currents lead to
different  angular distributions; and   
different ``cocktails'' of these currents may nevertheless result in compatible angular distributions.\newline 
The different theoretical models describing vector meson production 
should be confronted with the data presented here. 
This, although the data basis is still far from complete, may help to  
establish the reaction mechanism(s) of vector meson production.

\section*{Summary}
The cross sections presented in this paper extend and improve the experimental data base 
for the reaction $pp\to pp\omega$ at $\epsilon=93\, MeV$.
This results in an improved value of $\mathcal{R}_{\phi/\omega}$, 
which now is not only significantly larger than the na\"ive OZI value but also exceeds the more 
sophisticated predictions made by several theoretical approaches. 
The results of the comparison of all three angular distributions for $\phi$ and $\omega$ production are 
consistent with an identical production mechanism for both vector mesons.
This means that for the ``OZI-violation'' the main assumption of identical reaction processes is not disproved.  \newline
The angular distributions indicate the dominance of S-waves in the production mechanism. 
However, the differential cross section
 as function of the helicity angle
clearly shows the onset of higher partial waves in both $\omega$ and $\phi$ production.
Therefore, future experiments addressing the question of the reaction mechanism should concentrate on collecting 
data at excess energies around, and above, $100\,MeV$. 
In addition, the measurement of polarization observables is desired as it 
would help to develop a clearer picture of the reaction dynamics of $\omega$ and $\phi$ production
in proton-proton collisions.

\section*{Acknowledgment}
The authors would like to express their gratitude to the COSY staff for the operation 
of the accelerator during the experiment. Discussions with M.~Hartmann, A.~Sibirtsev, and C.~Hanhart are gratefully acknowledged.
This work was supported by BMBF and FZ-J\"ulich.



\newcommand{\btitle}[1]{\textit{``#1''},}
\newcommand{\bmag}[4]{#1~\textbf{#2}, #3 (#4).}
\newcommand{\ball}[5]{~\btitle{#1}~\bmag{#2}{#3}{#4}{#5}}
\newcommand{\ballhere}[5]{$\;$\bmag{#2}{ #3}{#4}{#5}}
\newcommand{\etall}[0]{\textit{et al.},}
\newcommand{\url}[1]{\mbox{URL: {\it #1}}}

\end{document}

%% file: ppomega_mschulte_COSY-TOF_2006_Tab1.tex
\small
\begin{table}[ttt]
\begin{center}
\begin{tabular}{| c || c | c ||  c | c | } \hline
 \multicolumn{1}{ c }{} 			 & \multicolumn{2}{ c }{ TOF  } & \multicolumn{2}{ c }{DISTO$\times 47.4$} \\ \hline 
 independent variable  &							  $a_0\;[\mu b/sr]$ 		   & $a_2\;[\mu b/sr]$  			   & $a_0\;[\mu b/sr]$  			   & $a_2\;[\mu b/sr]$   \\ \hline \hline 
\makebox[2.7cm][l]{proton CMS angle} & \makebox[2.0cm][c]{$0.71\pm0.03$} & \makebox[2.0cm][r]{$-0.13\pm0.08$} & \makebox[2.0cm][c]{----} & \makebox[2.0cm][c]{----} \\ 
\makebox[2.7cm][l]{meson CMS angle } & \makebox[2.0cm][c]{$0.68\pm0.04$} & \makebox[2.0cm][r]{$-0.01\pm0.12$} & \makebox[2.0cm][c]{$0.72\pm0.03$} & \makebox[2.0cm][r]{$ 0.06\pm0.06$} \\ 
\makebox[2.7cm][l]{helicity angle  } & \makebox[2.0cm][c]{$0.73\pm0.03$} & \makebox[2.0cm][r]{$ 0.42\pm0.07$} & \makebox[2.0cm][c]{$0.71\pm0.05$} & \makebox[2.0cm][r]{$ 0.26\pm0.12$} \\ 
\makebox[2.7cm][l]{Jackson angle   } & \makebox[2.0cm][c]{$0.75\pm0.03$} & \makebox[2.0cm][r]{$-0.09\pm0.07$} & \makebox[2.0cm][c]{$0.71\pm0.03$} & \makebox[2.0cm][r]{$-0.04\pm0.07$} \\ \hline
\end{tabular}
\vspace*{0.3 cm}
\caption{\label{tab:legendreresultsV2} \it Results of a Legendre fit to the data of TOF and DISTO presented in Fig.~\ref{fig:ergebnisse} 
(upper left to lower right). 
 Only the coefficients $a_0$ and $a_2$ are considered. 
(The DISTO data are scaled by the cross section ratio of  $\sigma_{\omega}/\sigma_{\phi}=47.4$ in order to facilitate a direct comparison.)
}
\end{center}
\end{table}
\normalsize